\documentclass{aa}  
\usepackage{natbib}
\usepackage[pdftex]{graphics}
\usepackage{graphicx}
\bibpunct{(}{)}{;}{a}{}{,}
\newcommand{\arcs}{$^{\prime\prime}$}
\usepackage{txfonts}
\begin{document}
   \title{Outflow or galactic wind: The fate of ionized gas in the halos 
of dwarf galaxies}


   \author{J. van Eymeren
          \inst{1}
          \and
          D.~J. Bomans\inst{1}
	  \and
	  K. Weis\inst{1,2}
	  \and
	  R.--J. Dettmar\inst{1}
          }


   \institute{Astronomisches Institut der Ruhr-Universit\"at Bochum,
              Universit\"atsstra{\ss}e 150, D-44780 Bochum\\
              \email{jeymeren@astro.ruhr-uni-bochum.de}\and Lise Meitner 
Fellowship
             }

   \date{Received date; accepted date}

 
  \abstract
   {H$\alpha$ images of star bursting irregular galaxies reveal a large amount
  of extended ionized gas structures, in some cases at kpc-distance away from
  any place of current star forming activity. A
  kinematic analysis of especially the faint structures in the halo of dwarf
  galaxies allows insights into the properties and the origin of this gas
  component. This is important for the chemical evolution of galaxies, the
  enrichment of the intergalactic medium, and for the understanding of the
  formation of galaxies in the early universe.}
   {We want to investigate whether the ionized gas detected in two irregular
  dwarf galaxies (NGC\,2366 and NGC\,4861) stays gravitationally bound to the
  host galaxy or can escape from it by becoming a freely flowing
  wind.}
   {
  Very deep H$\alpha$ images of NGC\,2366 and NGC\,4861 were obtained to
  detect and catalog both small and large scale ionized gas structures down to
  very low surface brightnesses. Subsequently, high-resolution long-slit
  echelle spectroscopy of the H$\alpha$ line was performed for a
  detailed kinematic analysis of the most prominent filaments and
  shells. To calculate the escape velocity of both galaxies and to
  compare it with the derived expansion velocities of the detected filaments
  and shells, we used dark matter halo models.}
   {We detected a huge amount of both small scale (up to a few hundred pc) and
  large scale (about 1--2\,kpc of diameter or length) ionized gas structures
  on our H$\alpha$ images. Many of the fainter ones are new detections. The
  echelle spectra reveal outflows and expanding bubbles/shells with velocities
  between 20 and 110\,km/s. Several of these structures are in accordance with
  filaments in the H$\alpha$ images. A comparison with the escape
  velocities of the galaxies derived from the NFW dark matter halo model shows
  that all gas features stay gravitationally bound.}
   {}

   \maketitle
%

\section{Introduction}

Irregular dwarf galaxies can be the sites of giant star formation
regions. The interplay between massive stars and the interstellar medium 
(ISM) has a large effect on the formation and the evolution of galaxies. 
Thereby, dwarf galaxies provide a perfect environment for this 
interaction as they are simple systems, fragile and hence likely to be
strongly affected by both external and internal processes 
\citep[e.g.,][]{Gallagher1984}. They generally have low metalicities, which 
may be a result
of their inability to retain newly synthesized metals. Additionally, their
surface brightness is very low, which could be due to an expansion of the
whole galaxy following the loss of a substantial fraction of its mass or due
to the cessation of star formation resulting from the loss of its ISM.\\ 
Numerous ionized gas structures up to kpc-size in and around the 
galactic plane of dwarf galaxies were found \citep[e.g.,][]{Bomans1997, 
Hunter1997, Martin1998, Bomans2001}. These structures can be divided into 
long, narrow filaments and ring-like structures. Using the definition of 
\citet{Bomans1997}, we refer to all ring-like structures with radii 
smaller than 500\,pc as superbubbles (SB). All ring-like structures larger 
than 500\,pc are called supergiant shells (SGS). Some of these gas features 
could be the relicts of former shell structures. Others mark the 
edges of shells that were produced by stellar winds and supernova 
explosions. Generally, these structures enclose large concentrations of 
massive stars, so-called OB associations. In this case, the ionization 
mechanisms are relatively well understood: Kinetic energy and momentum are
delivered by massive stars to their surroundings through stellar winds and
supernova explosions. A hot superbubble then expands into the ISM and sweeps
up the ambient gas which forms a thin, dense shell detectable in optical
emission line images. Having a sufficiently energetic and long-lasting
starburst, this shell can fragment, which allows the gas of the hot bubble to
escape.\\
However, ionized gas structures also exist at kpc-distances away from any
place of current star formation or hot massive stars \citep[e.g.,][]
{Hunter1993}. In this 
case, the ionization mechanisms are not obvious. Shock waves that are 
driven by a concentration of massive stars may sweep the interstellar gas 
out of the star forming region, which leads to the formation of a cavity. 
Due to lower 
densities than usual in the ISM, the photons can travel larger distances 
and can ionize much more distant neutral gas \citep[e.g.,][]{Hunter1997}. 
Apart from photoionization and shock ionization, turbulent mixing layers 
\citep{Slavin1993} and magnetic reconnection \citep{Birk1998} are additional, 
possible excitation mechanisms.\\
In theoretical models, the gas, most likely driven by collective supernovae, is
expelled into the halos of the galaxies. \citet{Norman1989} developed a
theory in which the gas is transported through tunnel-like features into
the halo, called chimneys. Depending on the strength of the gravitational
potential, the gas may be able to fall back onto the galactic disk, which is
described in the galactic fountain scenario \citep{Shapiro1976}.\\
All these theories are based upon models which try to explain the
observations. Several studies also show blowout scenarios by using numerical
simulations. \citet{MacLow1999} developed hydrodynamic models of dwarf 
galaxies by varying the energy input, the mass of the galaxy and the
metallicity. They address analytically and numerically the questions how
supernova explosions effect the interstellar medium of dwarf galaxies and what
happens to the gas, particularly to the metals. Their simulations show that 
only in low mass galaxies (${\rm \sim 10^{6}\,M_{\odot}}$) the 
probability of gas being able to leave the gravitational potential
increases. \citet{Silich2001} came to very similar results in their numerical
experiments and analytical estimates.\\
While hunting for ionized gas structures, the vicinity of the Giant
Extragalactic \ion{H}{ii} Regions (GEHR) is of huge interest. GEHRs exceed 
normal \ion{H}{ii} regions in size, luminosity and velocity dispersion. 
Usually, they harbor several concentrations of massive stars. Therefore, 
a lot of excitation should take place in and around these regions during the
lifetime of the OB associations. Our results will show that the presence of
excited gas around GEHRs does not necessarily correlate with the age of the OB
assocations and is sometimes not limited to the existence of OB
associations.\\
Most of the ionized gas structures seem to expand from their place of 
birth into the ISM. As the relatively low escape velocity of the dwarfs will
facilitate the removal of substantial amounts of interstellar matter, the 
question comes up whether these gaseous features stay gravitationally bound to 
the galactic disk (outflow) or whether they can escape from the gravitational 
potential by becoming a freely flowing wind (galactic wind). This is of 
special importance for the chemical evolution of galaxies, the enrichment of 
the intergalactic medium (IGM), and for the understanding of the formation of 
galaxies in the early universe \citep[e.g.,][]{Recchi2004}. The relative 
velocities of the ionized structures within about 1\,kpc around the star 
forming regions are quite low \citep[e.g.,][]{Martin1998}. Therefore, the gas 
appears to stay gravitationally bound to the host galaxies. No convincing 
case for a galactic wind has been found in a dwarf galaxy up to now 
\citep{Bomans2005}. Nevertheless, galactic winds are generally
regarded as a necessary ingredient to chemical and chemodynamical models 
of dwarf galaxies \citep[e.g.,][]{Hensler2004}. This apparent 
contradiction is most probably due to the previous inability to detect 
the faintest filaments at large distances from the host galaxies. The 
fastest moving shells most likely have the lowest densities, which 
corresponds to very low surface brightness in H$\alpha$.\\
 \begin{table}
      \caption[]{The sample of galaxies}
         \label{Sample}
	 $$
	   \begin{tabular}{lccc}
            \hline
            \noalign{\smallskip}
	    Parameters [unit] & NGC\,2366 & NGC\,4861 & References\\
	    \hline
	    \noalign{\smallskip}
	    Hubble Type$^{\mathrm{a}}$ & IB(s)m & SB(s)m &\\
	    $\rm m_{B}$ [mag] & $-$16.63 & $-$16.62 & (1)\\
	    $\rm D$ [Mpc] & 3.44 & 7.5 & (2), (3)\\
	    $\rm v_{sys}^{\mathrm{a}}$ [km/s] & 80 & 833 &\\
	    $\rm v_{rot,\,\ion{H}{i}}$ [km/s] & 67 & 54 & (4)\\
	    $\rm <\sigma>_{\ion{H}{i}}$ [km/s] & 7.7 & 8.4 & (4)\\
	    $\rm \sigma_{Peak,\,\ion{H}{i}}$ [km/s] & 14.3 & 19 & (4)\\
	    $\rm i$ [$\degr$] & 59 & 82 & (4)\\
	    $\rm M_{\ion{H}{i}}\left[10^9\,M_{\odot}\right]$ & 
	    0.57 & 1.14 & (4)\\
            \noalign{\smallskip}

            \hline
	   \end{tabular}
     $$ 
\begin{list}{}{}
\item[$^{\mathrm{a}}$] Data from NED
\end{list}
References. (1) \citet{Bomans2001}; (2) \citet{Tolstoy1995}; (3)
\citet{deVaucouleurs1991}; (4) \citet{Thuan2004}
   \end{table}
We examined ionized gas structures around two dwarf galaxies which are
very similar in mass, luminosity and shape. Very 
deep H$\alpha$ images show previously undocumented ionized structures, 
some of them with sizes of several kpcs, located in the halos 
\citep[e.g.,][]{vanEymeren2005}. Furthermore, we performed high-resolution
long-slit echelle spectroscopy covering several of the identified filaments in
order to measure their gas kinematics. The most relevant parameters of the
galaxies are listed in Table \ref{Sample}.\\
\begin{figure*}
\centering
\includegraphics[width=17.5cm]{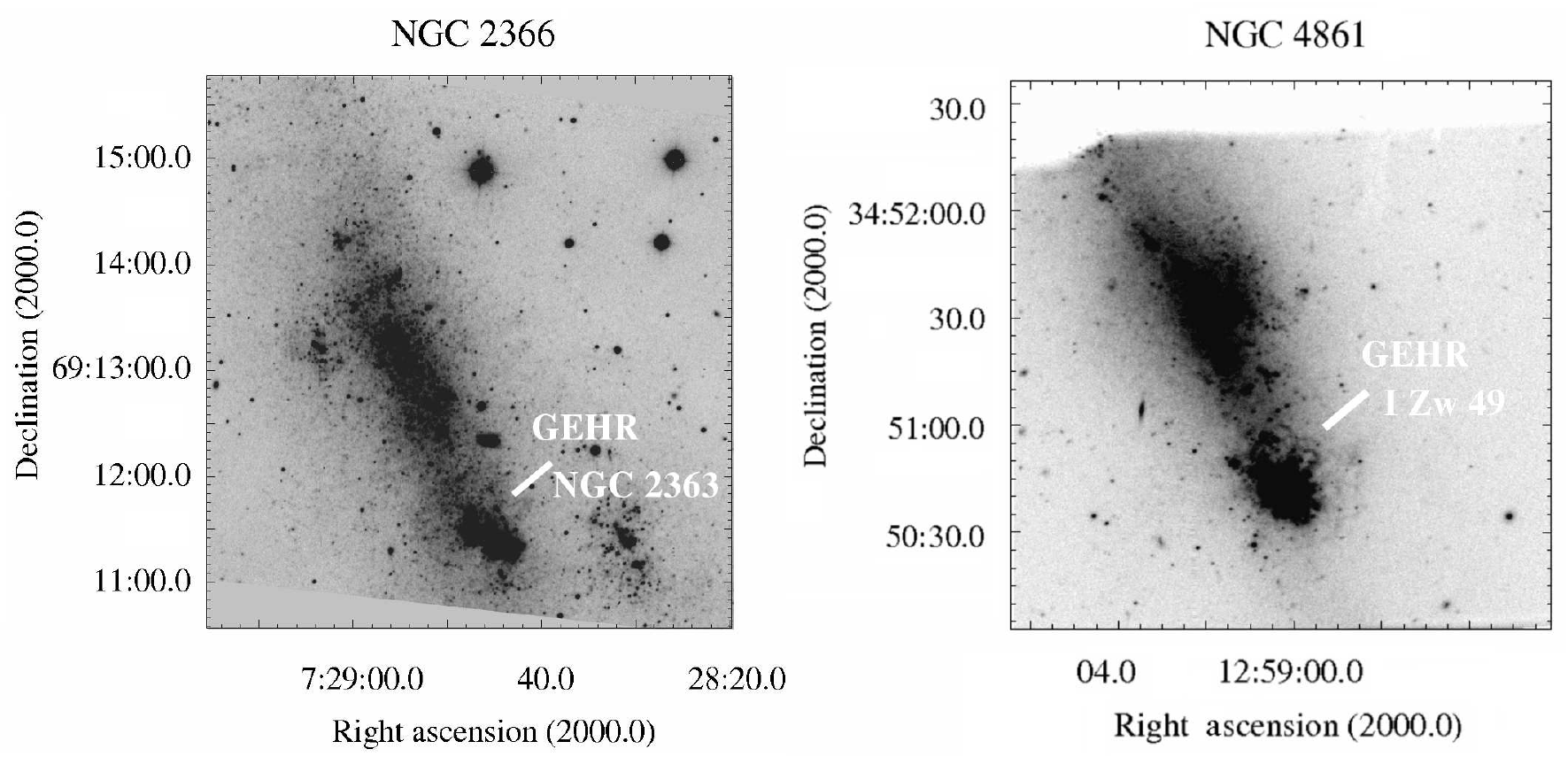}
\caption{R band images of NGC\,2366 (on the left) and NGC\,4861 (on the
  right). The GEHRs are marked in white. The white shadow on the top of
  NGC\,4861 was caused by the pick up arm of the guider tracking a bright star
  in the northern part of the galaxy and also affects the H$\alpha$ image (see
  Fig.~\ref{FigNGC4861}). Therefore, the northern-most part of the galaxy is
  missing. Nevertheless, it is the deepest R band exposure of NGC\,4861
  available so far.}
\label{Rimages}
\end{figure*}
  The distance of NGC\,2366 and its position on the sky place this galaxy
  into the M\,81\,group. Nevertheless, it can be regarded as an isolated
  galaxy. Its appearance in H$\alpha$ is dominated by the GEHR NGC\,2363 in
  the south-western end of the galaxy. This GEHR harbors two large OB
  associations, one in the core with an age of 1\,Myr and one in the eastern
  part of the core with an age of $\rm 3-5$\,Myr
  \citep{Drissen2001}. NGC\,4861 is very similar to NGC\,2366 concerning its
  shape. It is also dominated by a GEHR in the south-west, I\,Zw\,49, where
  most of the star formation occurs. A chain of small \ion{H}{ii} regions
  extends to the north-east. Although the galaxy shows almost no evidence for
  the existence of spiral structures \citep{Wilcots1996}, it is classified as
  SB(s)m. Its distance is more than twice the distance of NGC\,2366 (see Table
  \ref{Sample}). Figure~\ref{Rimages} shows our R band images of both
  galaxies.
\\
This paper is organized as follows. The observations and the data 
reduction are described in \S$\,$2. Sect.~3 presents the results of the
imaging data. The catalog of the H$\alpha$ filaments and shells can be found
in Appendix\,\ref{Appendix A}. In \S$\,$4, the analysis of the 
echelle spectra including the detection of expanding material follows. A
discussion of the results is presented in Sect.~5. \S$\,$6 summerizes the main
results.

\section{Observation and Data Reduction}
\subsection{Optical Imaging}
\label{2.1}
H$\alpha$ narrowband CCD images of the two irregular dwarf galaxies
were required to establish a catalog of the H$\alpha$ structures and to 
relate the slit positions of the spectra to the galaxies. Therefore, we 
used our image from the 3.5\,m telescope of the Calar Alto Observatory 
(NGC\,2366)\footnote{Based on observations collected at the Centro 
Astron$\rm\acute{o}$mico Hispano Alem$\rm\acute{a}$n (CAHA) at Calar Alto.}
and archival data from the 
3.6\,m Canada France Hawaii Telescope (NGC\,4861)\footnote{Guest User, 
Canadian Astronomy Data Center, which is operated by the Dominion
Astrophysical Observatory for the National Research Council of Canada's 
Herzberg Institute of Astrophysics.}, and from the HST (NGC\,2363). After the 
standard data reduction with the software package IRAF, the corresponding 
continuum images were subtracted to produce images of the H$\alpha$ line 
emission. In the case of the HST image of NGC\,2363, no continuum-subtraction
was done because no red continuum image was available. The image was
flux-calibrated by using the photflam value for the H$\alpha$
filter\footnote{WFPC2 Data Handbook chapter 5, URL:
  http://www.stsci.edu/instruments/wfpc2}. To estimate the contribution of the
continuum and to define an error for the energy calculation in
Sect.~\ref{5.1.1}, we took a different continuum filter F547M, that was
observed together with the F656N, scaled the flux of the stars to the stars in
the H$\alpha$ image and subtracted it from H$\alpha$. A comparison of the flux
of the faint H$\alpha$ filaments before the continuum-subtraction and
afterwards shows that the flux is roughly 6\% lower after the subtraction. As
we expect that a continuum subtraction with a red filter is much more
efficient than the one we did with F547M, the 6\% give us an upper limit for
the error. In Sect.~\ref{5.1.1}, we calculate the energy by taking into
account this uncertainty.\\
Finally, we used an adaptive filter which is based on the
H-transform \citep{Richter1991} to emphasize the weakest ionized gas 
features and to differentiate them from the noise.\\
Table \ref{Imaging} gives a short overview about the images.
 \begin{table*}
      \caption[]{Imaging -- some observational parameters}
         \label{Imaging}
     $$
         \begin{tabular}{lccccc}
            \hline
            \noalign{\smallskip}
	    Parameters [unit] & NGC\,2366 & NGC\,2363 & NGC\,4861\\
	    \noalign{\smallskip}
            \hline
            \noalign{\smallskip}
	    Date & 14.01.91 & 08.01.96 & 12.03.00\\
	    Telescope/Instrument & 3.5m Calar Alto Prime Focus & HST WFPC2 &
            3.6m CFHT OSIS\\ 
	    Filter & 658/10, Johnson\,R & F656N & 6570/48, 6493/1305\\
	    Exposure Time [s] & 1200, 200 & 1500 & 900, 300\\
	    Scale [arcsec/pix] & 0.38 & 0.1 & 0.15\\
	    FWHM [arcsec] & 1.1 & 0.2 & 0.8\\
            \noalign{\smallskip}
            \hline
         \end{tabular}
     $$ 
   \end{table*}
\subsection{Echelle spectroscopy of the H$\alpha$ line}
\begin{figure*}
\includegraphics[width=17.5cm]{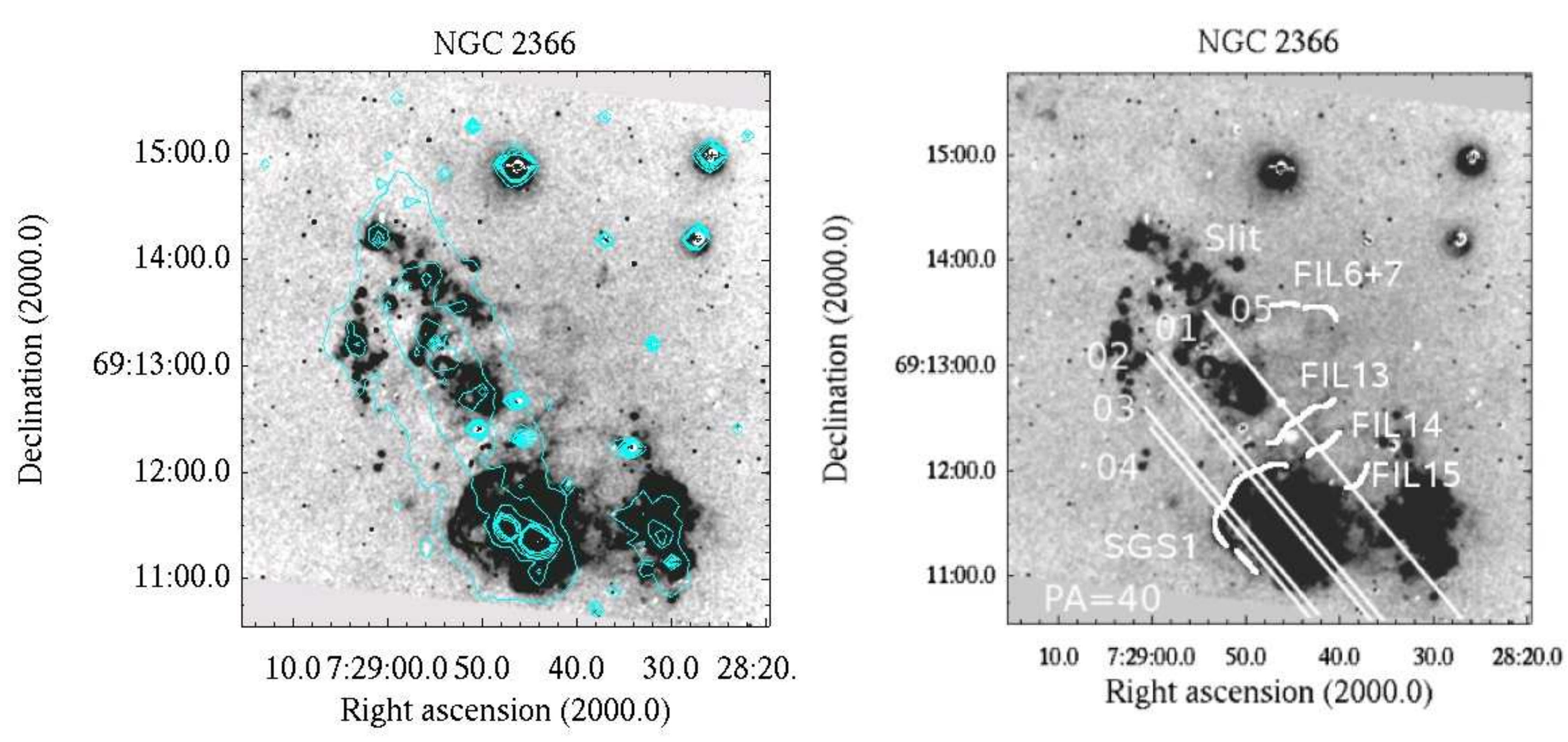}
\caption{NGC\,2366. Left panel: Continuum-subtracted H$\alpha$ image with the
  continuum contours overlaid in blue (grey). Right panel: Continuum-subtracted
  H$\alpha$ image with the five slit positions indicated by white lines and
  the most important ionized gas structures also indicated by white lines. The
  upper end of each line representing a slit position indicates the end of the
  true slit. The true lower end of these slits as well as upper and lower end
  of the slits in NGC\,4861 (see Fig.~\ref{FigNGC4861}) are located outside of
  the H$\alpha$ images.}
\label{FigNGC2366}
\end{figure*}

High-resolution long-slit echelle spectroscopy of both galaxies was 
performed with the 4\,m telescope of the Kitt Peak National 
Observatory from March 18th to 20th 1998. Inserting a post-slit 
H$\alpha$ filter with a width of 75\,{\AA} and replacing the cross
dispersion grating by a flat mirror, we selected the H$\alpha$ line at 
6563\,{\AA} and the two [NII] lines at 6548\,{\AA} and 6583\,{\AA}. We 
picked the 79 lines ${\rm mm^{-1}}$ echelle grating with a blaze angle of 
${\rm 63^{\circ}}$. The slit-width is about ${\rm 240\,{\mu}m}$ 
(corresponding to 1\farcs6), which leads to an instrumental FWHM at the 
H$\alpha$ line of about 13\,km/s.\\
All data were recorded with the long focus red camera and a 2048 x 2048 
Tek2 CCD. The pixel size is 0.08\,{\AA} along the dispersion and 0\farcs26
along the spatial axis. The slit-length was limited to 4\arcmin 46\arcs. The
seeing was about 1\arcs. For geometric distortion corrections we used star
spectra, for the wavelength calibration we used spectra of a Thorium-Argon
comparison lamp.\\
Close to the H$\alpha$ emission of both galaxies, we additionally detected
four night sky lines. In the spectra of NGC\,4861 they could be subtracted by
using the IRAF task \emph{background}. For this correction H$\alpha$ line-free
parts along the spatial axis are needed. In the spectra of NGC\,2366 the
H$\alpha$ emission is too extended to define an H$\alpha$ line-free area and
therefore to remove the night sky lines properly.  However, due to the
different redshifts of both galaxies, only the H$\alpha$ emission of NGC\,4861
is affected by one of these night sky lines. In the case of NGC\,2366 the
H$\alpha$ emission lies clearly separated between two night sky lines so that
this correction can be neglected.\\
The right panels of Figures~\ref{FigNGC2366} and \ref{FigNGC4861} show 
the slit positions of the spectra on the underlying continuum-subtracted 
H$\alpha$ image. We obtained five spectra of each galaxy with small offsets
from each other and with an exposure time
of 1800\,s (NGC\,4861 all slits, NGC\,2366 slit \emph{01} and \emph{02}) or
2400\,s (NGC\,2366 slit \emph{03}, \emph{04}, and \emph{05}). The position
angles were $\rm 40\degr$ for NGC\,2366 and $\rm 0\degr$ for NGC\,4861.\\
For the measurement of the emission lines the spectra were binned in the 
spatial direction by four pixels, which corresponds to about {1\arcs} matching
the seeing. At positions of very weak emission, we summed up over ten
pixels and used the IRAF task \emph{splot} in the interactive mode to
determine the peak wavelength and the Full Width at Half Maximum (FWHM). At
many locations, the emission line profile was double- or triple-peaked with
clear minima in the intensity between these peaks. We fitted such profiles
with two or three Gaussian components. The measured peak wavelengths were then
converted into heliocentric velocities.
\subsection{\ion{H}{i} data}
For a comparison of the measured H$\alpha$ velocities with rotation curves
derived from \ion{H}{i} data, we used published \ion{H}{i} moment maps by
\citet{Thuan2004} (their Fig.~1 and Fig.~7, bottom left panel). They used
archival VLA data with a spatial resolution of 12\farcs5 (NGC\,2366) and
15\farcs2 (NGC\,4861). We copied the velocity information for the
corresponding slit positions, which gives us a rough estimate for the
behaviour of the neutral gas in comparison to the ionized gas.
\section{Results -- The catalog of filaments}
Both galaxies show remarkable H$\alpha$ features partly of kpc-size. We divided
them into small scale (about a few hundred pc) and large scale structures
(about 1\,kpc and larger).\\
In the following subsections the structures of each galaxy are discussed. A
complete catalog of the H$\alpha$ structures can be found in
Appendix\,\ref{Appendix A}. All features were detectect by visual inspection
on our fully-reduced H$\alpha$ images (see Fig.~\ref{FigHafiln2366} and
Fig.~\ref{FigHafiln4861}) and afterwards measured manually. Only structures
above a 3$\sigma$ detection limit were considered for this analysis. To
measure the diameter of the superbubbles and supergiant shells, we began with
the intensity maximum on one side of the ring and ended on the intensity
maximum on the other side of the ring. The lengths of the filaments were
measured by starting from one end at an intensity larger than 3$\sigma$ and
stopping at the other end before the intensity drops below 3$\sigma$. The
errors for both measurements are about 0\farcs5 for each galaxy, which leads
to 8\,pc in the case of NGC\,2366 and and 17\,pc in the case of NGC\,4861. The
lower detection limit depends on the resolution and is 17\,pc in the case of
NGC\,2366 (FWHM of 1\farcs1, see Table~\ref{Imaging}) and 27\,pc in the case of
NGC\,4861 (FWHM of 0\farcs8).
\subsection{NGC\,2366}
\subsubsection{Small scale structures}
NGC\,2366 shows a wealth of small structures especially around the GEHR
NGC\,2363 and in the north-eastern part of its tail
(Fig.~\ref{FigHafiln2366}). These filaments have sizes of about a few hundred
pc (see Table \ref{Filsizea}). Most of them surround 
NGC\,2363 and seem to connect the GEHR to the small \ion{H}{ii} region in the
west. All filaments in the eastern part of NGC\,2363 are located at the inner
edge of the supergiant shell SGS1. Another important structure is the diffuse 
ionized gas in the north-western part of NGC\,2363. It is represented by a few 
small filaments (e.g., FIL14 and FIL15) which are all perpendicular to the 
major axis of NGC\,2366. FIL14 and FIL15 form the edges of an enormous outflow,
which is discussed in Sect.~4.1.1.\\
Altogether it seems that many of the smaller filaments connect the GEHR to the
large scale structures (e.g., SGS1) or to other \ion{H}{ii} regions.
\begin{figure*}
\centering
\includegraphics[width=17.5cm]{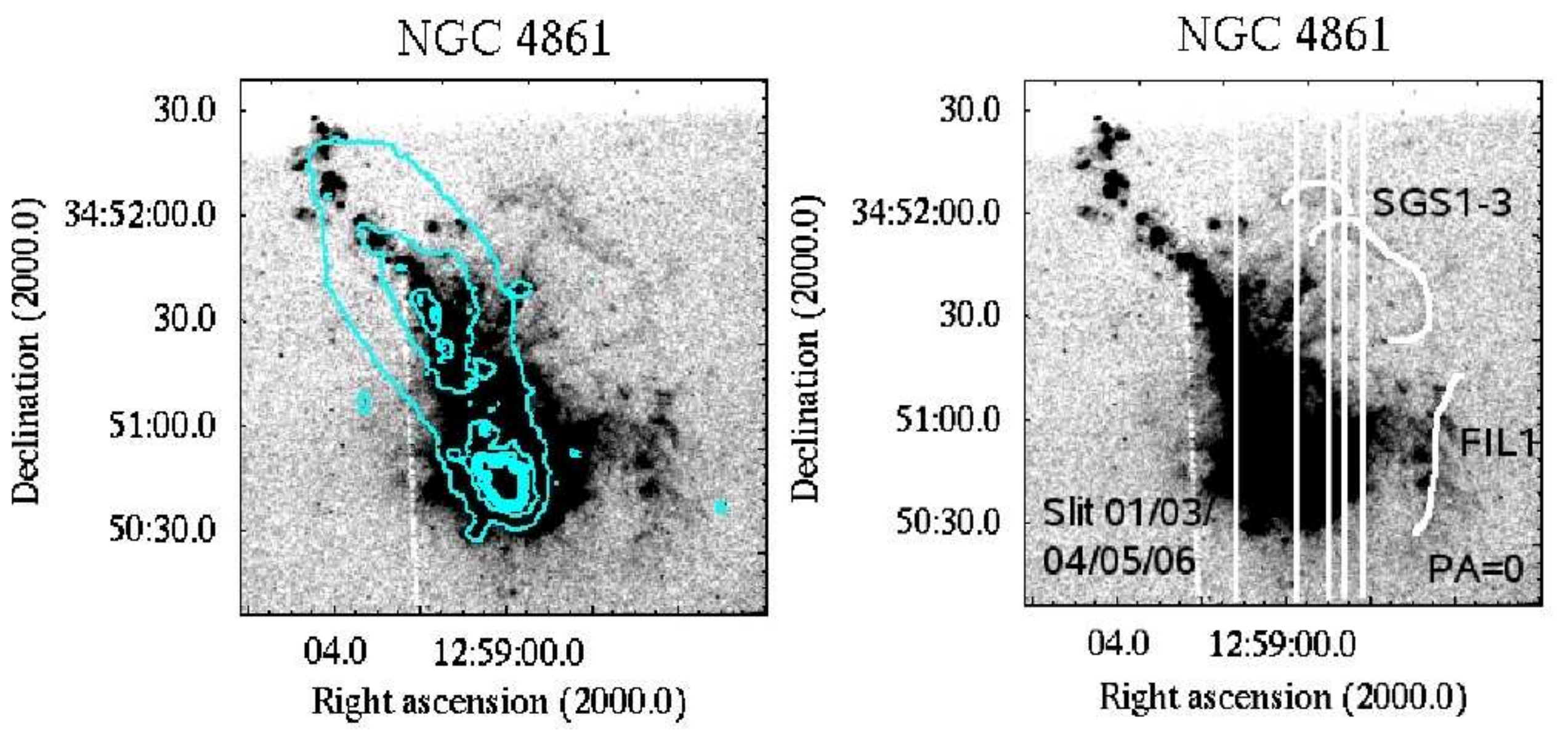}
\caption{NGC\,4861. Left panel: Continuum-subtracted H$\alpha$ image with
  the continuum contours overlaid in blue (grey). Right panel:
  Continuum-subtracted H$\alpha$ image with the five slit positions indicated
  by white lines (the numbers are rising from \emph{01} on the left to
  \emph{06} on the right side of the image) and with the largest structures
  marked in white. The quality of slit\,\emph{02} is too poor to give any
  reasonable results so that its position is left out on the right panel. The
  white shadow on the top was again caused by the pick up arm of the guiding
  camera (see Fig.~\ref{Rimages}) and hides the northern part of the galaxy.}
\label{FigNGC4861}
\end{figure*}
\subsubsection{Large scale structures}
The largest structure in NGC\,2366 is the above mentioned supergiant shell
SGS1 which was detected before by \citet{Bomans1992} --\,their shell 1\,--,
\citet{Hunter1993} --\,their features 1 and 2\,--, and \citet{Martin1998}
--\,her feature A\,--. It is located at the north-eastern part of the GEHR at
a distance of about 850\,pc from the center of NGC\,2363 and has a diameter of
about 900\,pc. It is connected to the GEHR via the smaller filaments as
discussed above. FIL6 and FIL7 together with FIL13 in the northern part of the
galaxy may form the edge of a giant shell which has fragmented as described in
Sect.~1. It then would have had a diameter of about 1\,kpc.
\subsection{NGC\,4861}
\subsubsection{Small scale structures}
Comparable to NGC\,2366, NGC\,4861 shows a high amount of smaller structures 
around its GEHR I\,Zw\,49 (Fig.~\ref{FigHafiln4861},
Table~\ref{Filsizeb}). But in this 
case, a lack of ionized filaments exists in the north-eastern part of the 
GEHR. In the western part, a complex web of filaments is visible on the 
H$\alpha$ image, which we detected as expanding material in our
spectra. Especially FIL9 to FIL11 at the western part of the tail, which
extend perpendicular to the major axis of NGC\,4861, seem to form a connection
to the large scale structure mentioned below.\\
The upper part of the tail of NGC\,4861 mainly consists of small \ion{H}{ii} 
regions and shows no diffuse filamentary emission. The white shadow on the top
of the image was caused by the pick up arm of the guiding camera tracking a
bright star in the northern part of the galaxy. This does not affect the
analysis because, as mentioned above, no ionized gas structures were found
around this area.
\subsubsection {Large scale structures}
The most prominent features in NGC\,4861 are the kpc-sized filaments 
extending from the north to the west of I\,Zw\,49 (SGS1-3, FIL1, see
Fig.~\ref{FigNGC4861} right panel). Assuming  that SGS1, SGS2 and SGS3 belong 
to one single shell, it would have a diameter larger than 2\,kpc.
\section{Results -- The kinematics of the ionized gas}
Figure~\ref{Figechspec} presents the echellograms
grouped by galaxy and arranged in a spatial sequence. A first glance reveals
a large amount of velocity structures. In a first step, we looked for
expanding gas and tried to connect the emission features to the cataloged
structures. Partially, a Doppler ellipse as described in \citet{Martin1998}
was detected, which gives evidence for an expanding shell structure. 
In some cases, we only detected emission at a constant wavelength which is 
Doppler-shifted with respect to the rest wavelength of the galaxy (measured in 
\ion{H}{i}). This indicates an expanding outflow. The following sections
present the results of our search.
\subsection{Detections of expanding material in the individual galaxies}
\subsubsection{NGC\,2366}
\begin{figure}
   \centering
   \includegraphics[width=.375\textwidth]{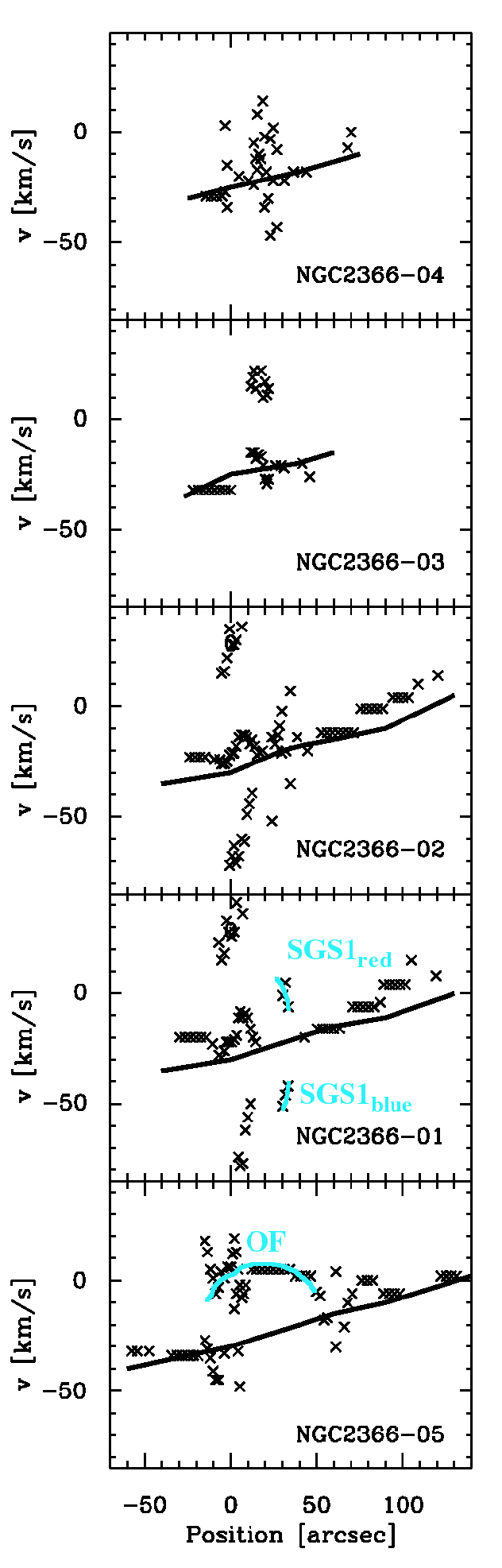}
        \caption{NGC 2366: Position-velocity diagrams of all slits arranged in
      a spatial sequence. The crosses mark the H$\alpha$ velocities, the solid 
      line represents the rotational motion of the galaxy derived from
      \ion{H}{i} maps by \citet{Thuan2004}. The most prominent ionized gas
      structures are marked in blue (grey). Referring to the position:
      Positive values go to the north-east. The measurement of the H$\alpha$
      peak velocity is very accurate and errors are only about few km/s,
      here indicated by the size of the crosses. The errors of the \ion{H}{i}
      velocity are smaller than 5\,km/s.}
         \label{FigNGC2366vp}
   \end{figure}

\begin{figure}
\centering
\includegraphics[width=.345\textwidth]{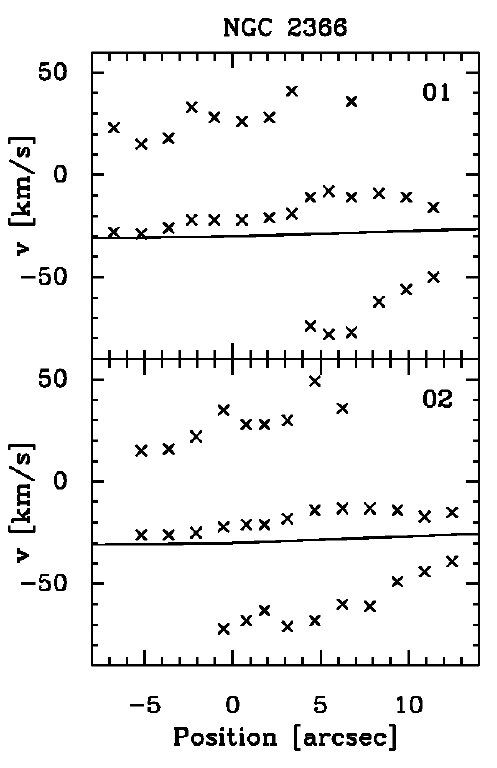}
\caption{Enlargement of the core region in slit\,\emph{01} and
  slit\,\emph{02}.}
\label{FigNGC2366pvenlarge}
\end{figure}

\begin{figure}
\centering
\includegraphics[width=.49\textwidth]{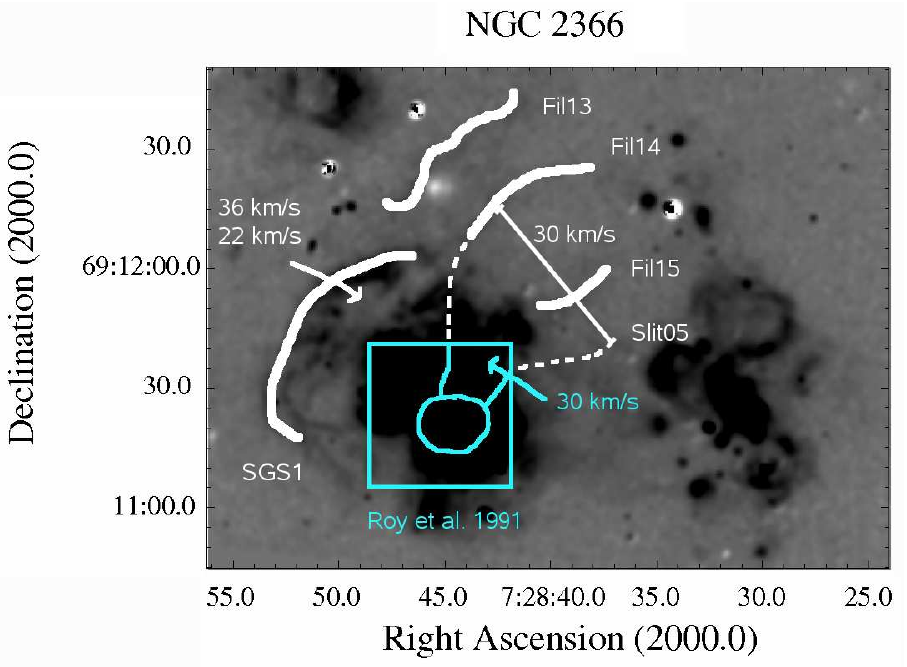}
\caption{Scheme of the north-western outflow of NGC\,2366. In blue (grey) the
  results of \citet{Roy1991}, in white our results. Slit\,05: Only the
  velocity offset of 30\,km/s is marked.}
\label{FigNGC2366scheme}
\end{figure}

Figure~\ref{FigNGC2366vp} shows the position-velocity (pv) diagram of the 
H$\alpha$ emission in NGC\,2366, Figure~\ref{FigNGC2366pvenlarge} shows an
enlargement of the core region in slits\,\emph{01} and \emph{02}. The solid
line marks the velocities of the \ion{H}{i} gas which are derived from the
\ion{H}{i} maps of \citet{Thuan2004} and which represent the circular velocity
of the galaxy (corrected for the redshift of the galaxy). The cross marks
define the heliocentric H$\alpha$ velocities corrected for the redshift of the
galaxy, which indicates the radial expansion velocities of the ionized gas 
compared to the \ion{H}{i} velocities. The location of the continuum emission
in slit\,\emph{01} indicating stars and therefore the center of the GEHR is
set to 0\arcs.\\
Generally, the velocity of the H$\alpha$ emission agrees with the velocity of 
the neutral hydrogen, which means that it takes part in the rotational motion 
of the galaxy. But at some positions there are significant deviations 
from the \ion{H}{i} velocities. Spectrum\,\emph{05} shows a blue-shifted 
component with a length of about 700\,pc and an expansion velocity of about
30\,km/s (marked as OF in Fig.~\ref{FigNGC2366vp}). As
mentioned above, this outflow is directly correlated with some of our detected
filaments on the H$\alpha$ image (especially with FIL14 and FIL15). Probably, 
there is a connection between the outflow and some detections 
of the H$\alpha$ emission leaving the GEHR NGC\,2363 to the north-west 
\citep{Roy1991, Martin1998}, which means that the ionized gas cone is larger 
than assumed before (see Fig.~\ref{FigNGC2366scheme}).\\
The supergiant shell SGS1 (see Sect.~3) is also detected in the pv
diagram. Unfortunately, the emission of the GEHR 
is so bright that it outshines all weaker emission features. Still, 
one can notice a small gap between the edge of the shell and the bright core 
of NGC\,2363 (see Fig.~\ref{FigNGC2366}). In this region the 
H$\alpha$ emission splits into two components, one blue-shifted and one 
red-shifted compared to the \ion{H}{i} velocity. From these measured 
velocities we estimate the expansion velocity of the shell, assuming that the
shell forms a hemisphere with a radius of 470\,pc and expands unequally. The
unequal expansion can be seen on the pv diagram of spectrum \emph{01} at
{30\arcs} (SGS1$\rm_{red}$ and SGS1$\rm_{blue}$). 
The red-shifted component shows lower velocities compared to the \ion{H}{i} 
data than the blue-shifted component. We derive as an upper limit 36\,km/s
blue-shifted and 22\,km/s red-shifted, which are very moderate expansion
velocities.\\
Apart from these striking features we generally see a lot of turbulence in
the core region of NGC\,2366. This is shown in all slits around position
0\arcs\, (Fig.~\ref{FigNGC2366vp}). The H$\alpha$ emission splits into three 
components, one following the \ion{H}{i} velocity, one blue-shifted and one
red-shifted. The enlargement of the core region of slit\,\emph{01} and
slit\,\emph{02} (see Fig.~\ref{FigNGC2366pvenlarge}) clearly shows the three
separate components.
\subsubsection{NGC\,4861}
\begin{figure}
   \centering
   \includegraphics[width=.4\textwidth]{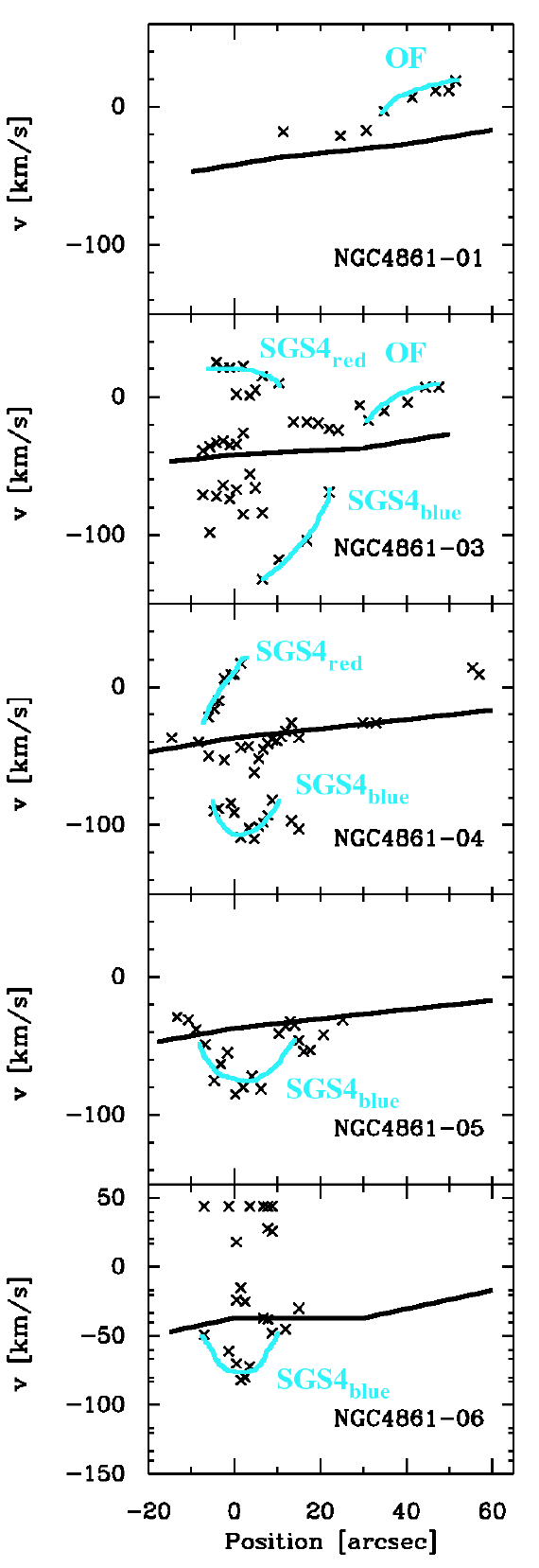}
      \caption{NGC 4861: Position-velocity diagrams of all slits arranged in a
      spatial sequence. The crosses
      mark the H$\alpha$ velocities, the solid line represents the rotational
      motion of the galaxy derived from \ion{H}{i} maps by
      \citet{Thuan2004}. The most prominent ionized gas structures are
      marked in blue (grey). Referring to the position: Positive values go to
      the north. The measurement of the H$\alpha$ peak velocity is very
      accurate and errors are only about few km/s, here indicated by the size
      of the crosses. The errors of the \ion{H}{i} velocity are smaller than
      5\,km/s.}
         \label{FigNGC4861vp}
   \end{figure}

NGC\,4861 shows a complex field of ionized gas structures which is very
similar to NGC\,2366 (see Sect.~3). Due to the limited throughput of the 
echelle spectrograph and unstable weather conditions during some parts of the 
observation, we only detect the emission of the luminous GEHR I\,Zw\,49. The 
faint and very interesting structures in the north-west of the galaxy (SGS1, 
SGS2, SGS3) are not visible in our spectra. Nevertheless, we find some 
striking features in the pv diagrams (Fig.~\ref{FigNGC4861vp}). In this case,
the location of the continuum emission in slit\,\emph{03} was set to
0\arcs. Slit\,\emph{02} was left out as the quality of the spectrum was too
poor to give any reasonable results.\\
In all spectra but spectrum\,\emph{01}, the H$\alpha$ emission splits into
several components. One component follows the rotational motion of the 
galaxy \citep[\ion{H}{i} data by ][]{Thuan2004}. Additionally, we find in all 
spectra but \emph{01} a blue-shifted component which partly has the 
appearance of a Doppler ellipse (SGS4$\rm_{blue}$). The radial expansion
velocity declines from 110\,km/s (spectrum\,\emph{03}) to 60\,km/s
(spectrum\,\emph{06}). In spectra\,\emph{03} and \emph{04}, there is also a
red-shifted component with an expansion velocity of 40\,km/s and 30\,km/s
respectively (SGS4$\rm_{red}$). The blue-shifted component may be a part of an
expanding shell with an expansion velocity of at least 110\,km/s. The
expanding material corresponds to the web of gaseous filaments in the western
part of I\,Zw\,49 (Fig.~\ref{FigHafiln4861}). It is limited by FIL1 which
could represent the  outer edge of the expanding shell. The red-shifted
component could belong to the same shell, but to verify this presumption, we
need deeper spectra of the whole galaxy. As the GEHR outshines all additional
emission line components, this shell cannot be detected on the H$\alpha$
image. Still, it is added with a blue- and a red-shifted part to the list of
ionized gas structures in Table~\ref{Filsizeb} as SGS4. Its diameter was
estimated from slit~\emph{03}.\\
The western outflow mentioned above was also found by \citet{Martin1998}. She
even detected a bipolar outflow going to the east and west of the galaxy. We
cannot verify the eastern outflow as we have no slit located at this
position. Instead, we also detect some outflowing gas in the northern part of 
the GEHR (see Fig.~\ref{FigNGC4861vp}, slit\,\emph{01} and slit\,\emph{03} OF),
which is described by \citet{Martin1998} as a faint red wing.
\section{Discussion}
\subsection{A complex web of filaments and their kinematics}
We found ionized shells and filaments in both galaxies, whereby the amount of
shell-like structures in comparison to the amount of filaments is very small,
although some of the filaments could be the relicts of former shell 
structures.\\
We divided all ionized gas features into
small and large scale structures. Many small scale structures are larger than
300\,pc, and we detected several large scale structures up to a size of 2
kpc. In many
cases it seems that in both galaxies the smaller filaments connect the
prominent \ion{H}{ii} region to the large scale structures or to smaller 
\ion{H}{ii} regions as in the case of NGC\,2366. Both galaxies contain some
kpc-sized filaments which are parallel to their major axes (e.g., SGS1-3 in
NGC\,4861). \ion{H}{i} intensity maps \citep{Thuan2004} show that the neutral
gas is much more extended than the ionized gas. That implies that the ionized
shells are running into an extended \ion{H}{i} envelope, which leads to some
interaction between both phases. Unfortunately, our echelle spectra do not
give any kinematic information about those shells so that we cannot make any
statements on which processes happen in the halo. Deeper spectra and
Fabry-P$\rm\acute{e}$rot interferometry are proposed to improve the
sensitivity of our observations.\\ 
Kpc-scale expanding shells which are elongated in the general direction
  of the \ion{H}{i} minor axis were found before in amorphous dwarfs
\citep{Marlowe1995} and in I\,Zw\,18 \citep{Martin1996}. Furthermore, narrow
elongated filaments which seem to connect the disk with the halo were recently
detected in spiral galaxies by \citet{Rossa2004}.\\ 
Especially NGC\,4861 shows these huge supergiant shells (SGS1-3) and several
finger-like structures (FIL9-11) which emanate from the disk into the
halo and seem to connect the disk with the shells. The chimney scenario (see
Sect.~1) is one possible explanation. Another explanation are Rayleigh-Taylor
instabilities. If two neighboring gas layers of different density are
perturbed, potential energy is released in the sense that the heavier material
moves down under the gravitational field and the lighter material is displaced
upwards. That means in our case that under the assumption that SGS1-3 have
once formed a single shell, this shell has ruptured at some points and the gas
is now falling down. This downfalling gas has the appearance of a finger,
which is exactly what we see on the H$\alpha$ image.\\
Referring to these large scale structures, we have to check their expansion 
velocities by obtaining deeper spectra than the currently available echelle 
spectra. Only then we can make precise statements on their evolution. The
spectra of NGC\,2366 already indicate that the gas near the perpendicular
filaments expands from the GEHR, whereas we have no such information on
NGC\,4861.\\
Most of the smaller and disk-near expanding ionized gas structures which we
found in the echelle spectra 
can be correlated to the identified features in the H$\alpha$ images. Some of 
the features have the appearance of a Doppler ellipse in the pv diagram or they
form at least a part of a Doppler ellipse as can be seen in e.g.,
Figures~\ref{Figechspec} and \ref{FigNGC4861vp}, slit\,\emph{05} of
NGC\,4861. At many other positions, we just detect a constant, Doppler-shifted
component (e.g., Fig.~\ref{FigNGC2366vp} slit\,\emph{03}).\\
The expansion velocities are usually very moderate (up to 50\,km/s), but in 
NGC\,4861, we found an expanding shell with a velocity of about 110\,km/s.
\subsubsection{The origin of the expanding shells}
\label{5.1.1}
We want to study one of our detected shells in more detail: SGS1 in NGC\,2366
is found to expand unequally with very moderate velocities (see Sect.~4.1.1).
One reason for an unequally expanding shell are possible density 
inhomogeneities. Probably, the gas density behind the shell is higher than in 
front of it (along our line of sight). Thus, the ionized gas shell has to push 
more material if the density is higher, which leads to a deceleration of its 
expansion velocity. We also estimated the kinetic energy of this shell under 
the assumption of a very thin hemisphere of radius $\rm r=470\,pc$ and 
thickness $\rm \delta{r}=5\,pc$. The mass was calculated from the 
flux-calibrated HST image (see Sect.~\ref{2.1}). We get a kinetic energy
of $\rm E(H^{+})=4.6\cdot10^{50}\,ergs$. This is just an upper limit as we
overestimate the H$\alpha$ flux (see Sect.~\ref{2.1}). Taking into account the
error of the flux calibration, we get a value that is 3\% lower.\\
The age of expansion can be estimated (under the assumption of a
constant expansion) from $\rm t \approx \frac{R}{v_{exp}}$, which gives us
about $10^{7}$\,years. Comparing this to the ages of the two local star
clusters (1 and 5\,Myr old respectively) results in the conclusion that we do
need a former star formation event. Nevertheless, there must be another
excitation or ionization mechanism which keeps the shell expanding. This could
possibly be done by stellar winds from the two current star clusters.
\subsubsection{Large line widths in NGC\,2366 and NGC\,4861}
The shells and filaments and especially the GEHRs of both galaxies show a Full
Width at Half Maximum (FWHM) of about 30 to 50\,km/s (corrected for 
instrumental
broadening). This gives us a velocity dispersion of 11 to 19\,km/s. These
values are comparable to those of GEHRs \citep{Hunter1997}. But GEHRs harbor
large OB associations as energy source, whereas no such energy sources exist
around most of the detected filaments. \citet{Hunter1997} explain such a high
dispersion with a high dispersion of the \ion{H}{i} gas. Looking at the
velocity dispersion of the neutral hydrogen (see Table\,\ref{Sample}), the 
H$\alpha$ dispersion is higher than the average \ion{H}{i} dispersion, but
fits well the peak dispersion (NGC\,2366: 14.3\,km/s, NGC\,4861: 19\,km/s). A
comparison with the \ion{H}{i} peak dispersion is more reliable as we trace
H$\alpha$ only in the central regions of the galaxies. Therefore, our 
observations prove the statement of \citet{Hunter1997}.\\
The question is now why especially the GEHRs of both galaxies show such a high
FWHM (e.g., the GEHR of NGC\,2366 splits into three different components). A 
work by \citet{Yang1996} examined the most luminous \ion{H}{ii} region NGC\,604
in M\,33. They also found broadened H$\alpha$ lines with FWHMs of about
40\,km/s. If one assumes the thermal component of H$\alpha$ to be $\rm
~20\,km/s$, these high FWHMs cannot only be explained by thermal
broadening. Other mechanisms which are suggested and discussed by
\citet{Yang1996} are stellar winds, and supernova remnants. 
\subsection{Outflow or galactic wind?}
Our results of both H$\alpha$ imaging and echelle spectroscopy lead to the 
question of the fate of the expanding gas. To determine whether it can escape 
from the gravitational potential (galactic wind) or not (outflow), we compared 
its expansion velocity to the escape velocity calculated by using the dark 
matter halo model by \citet{Navarro1996}, for short NFW-model. In this model, 
the galaxy is dominated by a halo of dark matter, and the baryonic matter of 
the disk is neglected. The circular and the escape velocity were calculated by 
\begin{equation}
\label{vrot}
\rm v_{rot}=\sqrt{\rm \frac{G\,M_{s}}{3}\,\frac{r}{r+r_{s}}^{2}}
\end{equation}
and
\begin{equation}
\label{vesc}
\rm v_{esc}=\sqrt{\rm 2\,\left|-\frac{G\,M_{s}}{r}log\left(1+\frac{r}
{r_{s}}\right)\right|}
\end{equation}
We compiled the pv diagrams of NGC\,2366 and NGC\,4861, where position\,0
corresponds to the dynamic center of the galaxies and all velocities are
corrected for the redshift.\\
In the case of both galaxies, we used the \ion{H}{i} data by
\citet{Thuan2004}. By varying the reference mass ${\rm M_s}$ and the virial
radius ${\rm r_s}$, we calculated the rotation curves (using Eq.~\ref{vrot})
in order to get the best approximation to the \ion{H}{i} data. In
Figures~\ref{DMH1} and \ref{DMH2}, two different parameter sets (${\rm M_s}$
and ${\rm r_s}$) for each galaxy are shown. Using the same reference masses and
virial radii than for the rotation curves, we calculated the escape velocities
by Eq.~\ref{vesc}. Additionally, we plotted the expansion velocities of
our detected outflows at the corresponding distance from the dynamic center
of the galaxies.\\
Tables~\ref{Filsizea} and \ref{Filsizeb} show several
filaments and shells that could also be detected on the slit spectra. In
  most of the cases, one prominent structure is based on several smaller
  structures, e.g., in NGC\,2366 all filaments from FIL24 to FIL28 belong to
  the supergiant shell SGS1. Therefore, we only consider the most prominent
  structures for our analysis, i.e. the hemisphere limited by SGS1 (the radial
  velocities have been estimated in Sect.~4.1.1) and the north-western outflow
  marked by FIL14 and FIL15 in NGC\,2366 and the supergiant shell visible on
  the pv diagrams in NGC\,4861 (see Sect.~4.1.2). All other structures are not
  well-defined so that we cannot tell anything about their symmetry and
  therefore about their true expansion velocity.\\
As we assume a spherical symmetry for both expanding shells around the
  GEHRs of NGC\,2366 and NGC\,4861, the expansion velocities do not need to be
corrected for the inclination of the galaxies. At every position where we
intersect the shells we get the true expansion velocity. 
In the case of the outflow in NGC\,2366, we cannot define a geometry. But if
one assumes an inclination angle of 59\degr, measured from \ion{H}{i}
observations (see Table~\ref{Sample}), we get an expansion velocity of 35\,km/s
instead of our measured value of 30\,km/s, which gives a velocity increase of
only 17\,$\%$, still far below the derived escape velocity of
the galaxy.\\
The chosen size of the reference mass is in all cases by orders one or two
higher than the \ion{H}{i} mass derived from the \ion{H}{i} velocity maps 
(see Table \ref{Sample}). That means that the determination of the mass of the
galaxy from the \ion{H}{i} velocities is only a rough estimate.\\
A comparison of the \ion{H}{i} data with the rotation curves derived from 
Eq.~\ref{vrot} shows that neglecting the cores of dwarf galaxies 
leads to a discrepancy between the dark matter halo rotation curves and 
the observed \ion{H}{i} data at small distances from the dynamic center. That 
means that the baryonic matter in the disk has a significant influence on 
the rotation of the galaxy near the dynamic center. In the outer parts, 
dwarf galaxies are dominated by the dark matter halo.
\subsubsection{NGC 2366}
\begin{figure}
   \centering
   \includegraphics[width=.43\textwidth]{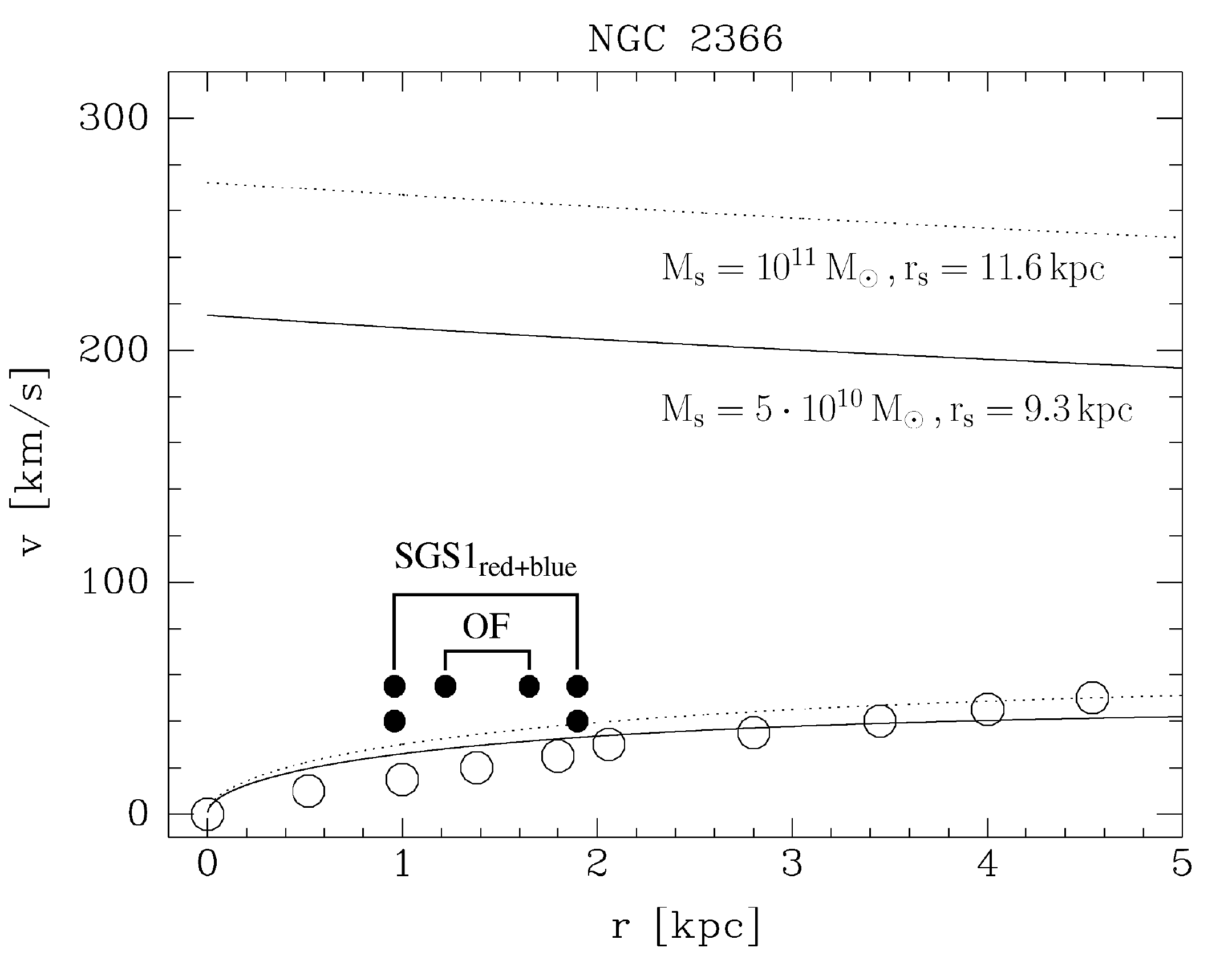}
      \caption{Position-velocity diagram of NGC\,2366. Different velocities
	are plotted versus the distance from the dynamic center of the
	galaxy. The solid and dashed lines represent two different parameter
	sets of the mass and the radius. The rotation curves follow more or
	less the \ion{H}{i} position-velocity data described in
	\citet{Thuan2004}, here drawn as open symbols. The lines in the upper
	part show the escape velocity. The expansion velocities of our
	detected shells and outflows are presented as solid symbols (beginning
	and end point of each structure). Referring to the
	velocity, the sizes of the solid and open symbols show the errors
	which are about 5\,km/s for measuring the expansion velocity and about
	10\,km/s for the \ion{H}{i} velocities. Positional errors are
	negligible. Therefore, they are not presented here.}
         \label{DMH1}
   \end{figure}

NGC\,2366 is dominated by the GEHR NGC\,2363. Several ionized structures 
emanate from this actively star-forming region. Both the expansion velocities
of the shell SGS1 and the outflow to the north-west of NGC\,2363 are shown 
in Figure~\ref{DMH1}. Plotted are the beginning and the end 
point of the structures corresponding to their distances from the dynamic 
center. SGS1 is presented with both expansion velocities. As we had to 
choose very high reference masses in comparison to the \ion{H}{i} mass of 
the galaxies to fit the rotation curve to the \ion{H}{i} data, we get very 
high escape velocities of about 200\,km/s and higher. The expansion velocities 
are moderate and remain far below the escape velocities. That means that the 
gas is still gravitationally bound to the galaxy.\\
One has to have in mind that at least the velocity of the western outflow is 
just a radial velocity. The real velocity will be higher so that our values 
just represent a lower limit of the expansion velocities.
\subsubsection{NGC 4861}
\begin{figure}
   \centering
   \includegraphics[width=.43\textwidth]{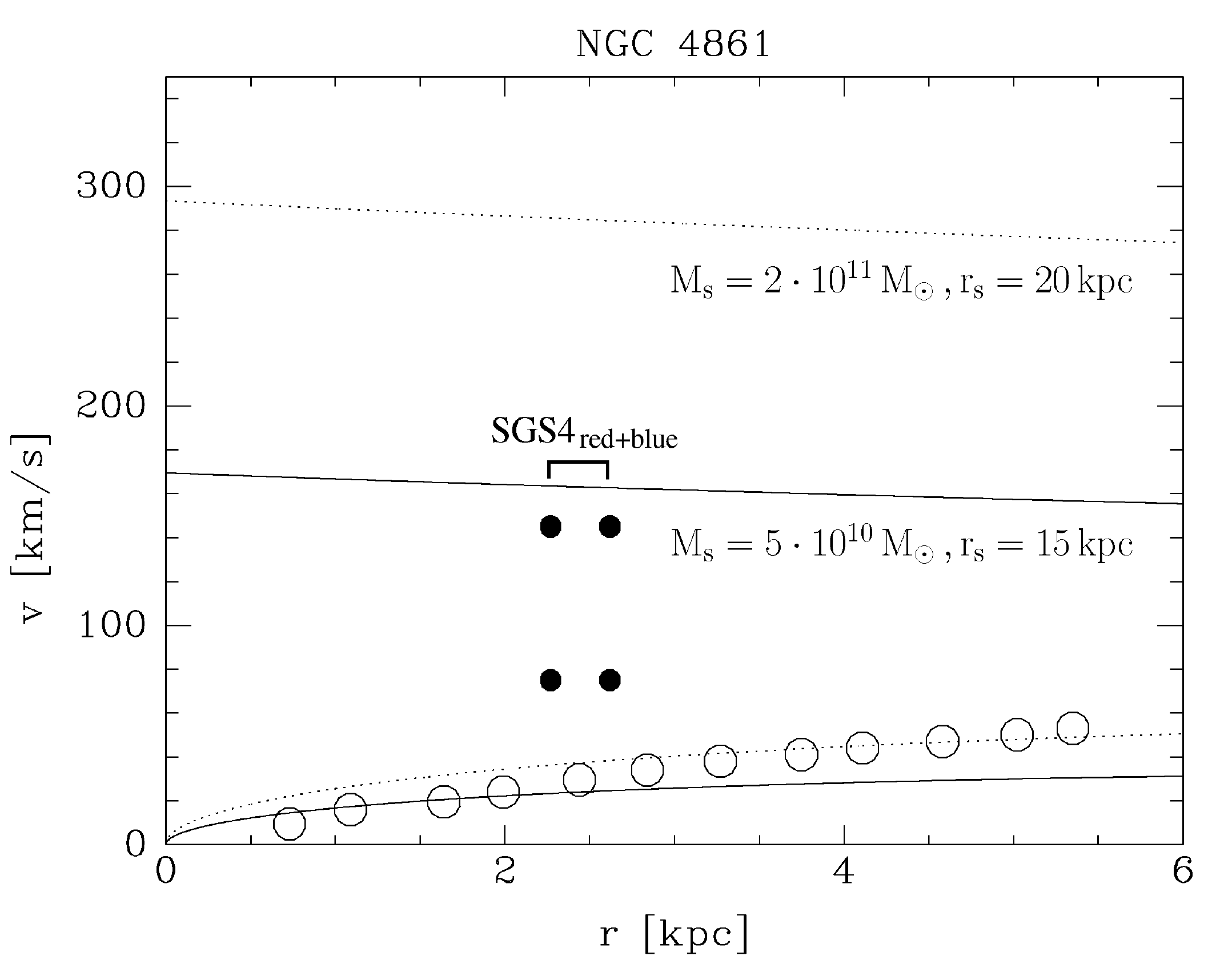}
      \caption{Position-velocity diagram of NGC\,4861. The same as in
      Figure~\ref{DMH1}. The \ion{H}{i} data are again taken from
      \citet{Thuan2004}.}
         \label{DMH2}
   \end{figure}

In NGC\,4861 (Fig.~\ref{DMH2}) we have a similar situation as in
NGC\,2366. We found one large outflow which probably forms an expanding
shell. The value of the red-shifted component is moderate ($<$ 40\,km/s) and
remains far below the escape velocity of the galaxy. The blue-shifted
component of the outflow expands with a much higher velocity (about
110\,km/s) and therefore nearly reaches the escape velocity, at least in the
case of the lower values for mass and virial radius. But when looking at the
corresponding rotation curve it becomes obvious that choosing a low mass and a
low virial radius gives us a very poor model to the \ion{H}{i} velocities of
\citet{Thuan2004}. From a  distance r of 2.5\,kpc on, which corresponds to the
position of the outflow, the \ion{H}{i} rotation curve and the model are not
in good agreement. Only by going to higher masses and higher virial radii,
this fit can be improved, which implies that the escape velocity is rising and
that therefore the faster expanding part of the shell definitely stays below
the escape velocity.
\subsection{Outflow!}
Both galaxies show outflows with expansion velocities between 20 and 
110\,km/s. Generally, these velocities stay below the escape velocities 
of the galaxies derived from the NFW-model.\\
A comparison of our results with the 1d chemodynamical models of
\citet{Hensler2004} shows that in the end the dark matter defines the fate of
the expanding gas. Their models predict that most of the gas
(galaxy mass of ${\rm \sim 10^{9}\,M_{\odot}}$) or even all the gas (galaxy 
mass of ${\rm \sim 10^{10}\,M_{\odot}}$) can leave the gravitational potential 
as a galactic wind by neglecting the dark matter. Therefore, no further star 
formation due to a new collapse 
of the gas is possible. However, the presence of dark matter increases the 
escape velocity of the galaxy. The gas cannot flow out that easily so
that the galaxy is enriched by metals and further star formation can take
place. This could be true for our sample galaxies. In both of them we detected 
gas which seems to flow out of the galactic disk, but stays closely to the 
galaxy (see Sect.~3). We measured the expansion velocities of some of these 
structures, which are relatively moderate and not sufficient for a galactic 
wind. A comparison with the models of \citet{MacLow1999} (Sect.~1) confirms
our results.\\ 
\citet{Martin1998} also used dark matter halo models to compare the escape
velocity of the galaxies to the expansion velocities of the shells. She showed 
that only the smallest galaxies of her sample, Sextans\,A and 
I\,Zw\,18, have shells with expansion velocities comparable to the escape
velocities. Otherwise, the expansion velocities lie far below the escape
velocities.\\
That means that we have to look at galaxies of very low mass when hunting for
galactic wind structures. On the other hand, it is not clear yet what happens
to the large scale structures at kpc distances from the galactic disk. Neither
\citet{Martin1998} nor we were able to detect them in the spectra.\\
Therefore, our next steps are further investigations of these two and other 
dwarf galaxies. We especially have to find a method to detect the weakest 
emission in the halos of the galaxies.
\section{Summary}
We examined two irregular dwarf galaxies which are very similar according to
their mass, shape and luminosity.\\
First, we used H$\alpha$ images to create a catalog of shells and filaments 
(see Appendix\,\ref{Appendix A}). We performed high-resolution long-slit
echelle spectroscopy of the most prominent emission features in order to
analyze their kinematics. Finally, we used dark matter halo models to get an
idea whether the expanding structures can leave the gravitational potential or
not.\\\\
In both galaxies we found both small scale (up to a few hundred pc) and large
scale (about 1--2\,kpc) ionized gas structures. The GEHRs are mainly
surrounded by smaller filaments which sometimes seem to connect the GEHRs to
the large scale filaments or to neighboring \ion{H}{ii} regions (e.g., in
NGC\,2366). Especially in the outer parts of the galaxies, the filaments have 
large scale sizes. They are located at distances up
to several kpc away from any place of current star formation. Thus, one has to 
think about the ionizing processes. Probably, the ionizing OB association has 
already died and the shells and filaments are only some relicts of a former 
star formation event. This would give us direct hints with respect to the age 
and the development of galaxies.\\
Furthermore, we need to explain the disrupted structure of the giant
shells and the connecting filaments between the disk and the halo. As we do
not have any kinematic information, we cannot prove whether these structures
can be caused by Rayleigh-Taylor instabilities, finger-like emanating gas
according to the chimney model or interaction with a surrounding \ion{H}{i}
envelope, just to mention a few possible scenarios. This has to be
investigated in more detail with much deeper data.\\
Both galaxies show outflowing material. The expansion velocity varies from 
20\,km/s to 110\,km/s. Using the dark matter halo model by 
\citet{Navarro1996} in order to compare the expansion velocities to the escape 
velocities of the galaxies, we found that in all cases the 
expansion velocity stays below the escape velocity. This result fits to the 
predictions of \citet{MacLow1999} and the studies of \citet{Martin1998}, but
it does not fit to the 1d chemodynamical models of \citet{Hensler2004}. 
It draws our attention to the faint large scale structures in the outer parts
of the galaxies and to mass-poor galaxies with a low gravitational potential.
\begin{acknowledgements}
The authors would like to thank U. Hopp for providing his image of
NGC\,2366. We thank Lutz Haberzettl, Chaitra Narayan, and Clemens Trachternach
for helpful comments and useful discussions.\\
This research made use of the NASA's Astrophysics Data
System Abstract Service, the LEDA database (http://leda.univ-lyon1.fr), and
the NASA/IPAC Extragalactic Database (NED) which is operated by the Jet 
Propulsion Laboratory, California Institute of Technology, under contract with 
NASA.
\end{acknowledgements}
\bibliographystyle{aa}
\bibliography{7335}
\appendix
\section{The H$\alpha$ images and catalogs}
\label{Appendix A}
Here, we present the continuum-subtracted H$\alpha$ images of NGC\,2366 and
NGC\,4861 with the ionized gas structures marked in white. Additionally, the 
identified gas features are listed in the following tables. We measured the
  angular sizes of the ring-like structures and the filaments and converted
  them to linear sizes assuming the distances given in Table~\ref{Sample}. For
  the detection method and the measurements of the structures see Sect.~3.\\
For all structures we could identify in our
spectra the values of the heliocentric velocity and the FWHM (corrected for
instrumental broadening) are also presented in Table~\ref{Filsizea} and
\ref{Filsizeb}.
\begin{figure*}[h]
\centering
\includegraphics[width=12cm]{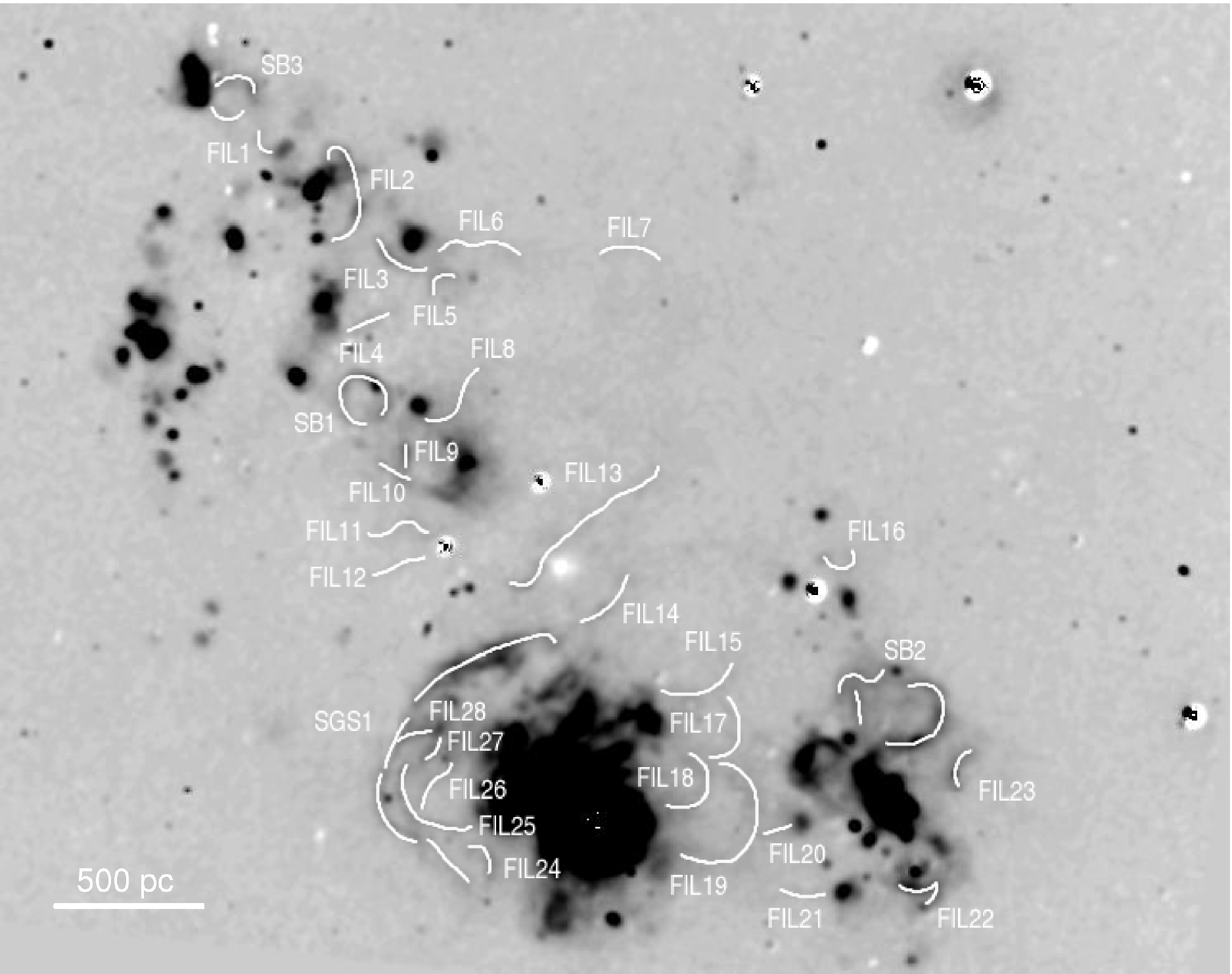}
\caption{Continuum-subtracted H$\alpha$ image of NGC\,2366: position and
  designation of the filaments and
  shells are superimposed in white for better visibility. The contrast was
  chosen in a way to demonstrate the small scale structures. For a
  presentation of the large scale filaments see Figure~\ref{FigNGC2366}. A
  measure in pc to estimate the size of the structures is given in the lower
  left corner of the image. The distance of NGC\,2366 is given in
  Table\,\ref{Sample}.}
\label{FigHafiln2366}
\end{figure*}
\begin{figure*}[h]
\centering
\includegraphics[width=12cm]{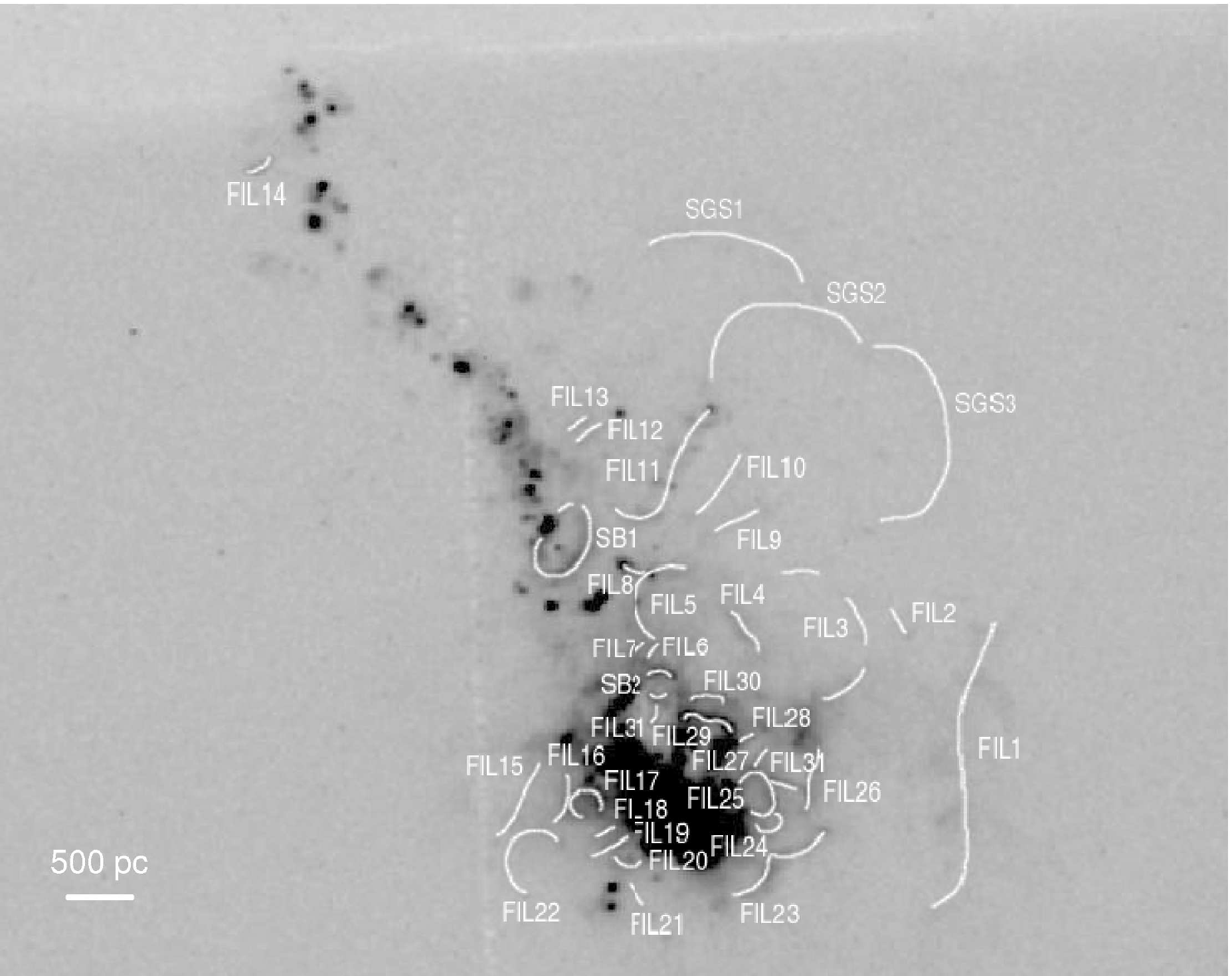}
\caption{Continuum-subtracted H$\alpha$ image of NGC\,4861: position and
  designation of the filaments and
  shells are superimposed in white for better visibility. Again we chose a
  contrast for stressing the small scale structures. The large scale structure
  is shown in Figure~\ref{FigNGC4861}. To estimate the size of the structures,
  a measure in pc is given in the lower left corner of the image. The distance
  of NGC\,4861 can be found in Table\,\ref{Sample}.}
\label{FigHafiln4861}
\end{figure*}
\\\\\\\\\\\\\\\\\\\\\\\\\\\\\\\\\\\\\\\\\\\\\\\\\\\\\\\\\\\\\\\\\\\\\\\\\\\\\\\\\\\\\\\\\\\\\\\\\\\\\\\\\\\\\\\\\\\\\\\\\\\\\\\\\\\\\\\\\\\\\\\\\\\\\\\\\\\\\\
 \begin{table}
      \caption[]{The most prominent structures and their sizes in
      NGC\,2366.}
         \label{Filsizea}
     $$
         \begin{tabular}{lcccc}
            \hline
	    \hline
	    \noalign{\smallskip}
            Source & Diameter & Diameter & $\rm v_{helio}$ & FWHM\\
	    & [\arcs] & [pc] & [km/s] & [km/s]\\
            \noalign{\smallskip}
            \hline
            \noalign{\smallskip}
            SGS1 & 59 & 940 & 49/99 & 26/38\\
	    SB1 & 11 & 180\\
            SB2 & 19 & 300\\
	    SB3 & 9 & 150\\
	    \noalign{\smallskip}
            \hline
	    \hline
	    \noalign{\smallskip}
	    \noalign{\smallskip}
	    \noalign{\smallskip}
	    \hline
	    \hline
	    \noalign{\smallskip}
            Source & Length & Length & $\rm v_{helio}$ & FWHM\\
	    & [\arcs] & [pc] & [km/s] & [km/s]\\
            \noalign{\smallskip}
            \hline
            \noalign{\smallskip}
	    FIL1 & 7 & 110\\
	    FIL2 & 32 & 510\\
	    FIL3 & 14 & 230\\
	    FIL4 & 9 & 140\\
	    FIL5 & 10 & 160\\
	    FIL6 & 23 & 370\\
	    FIL7 & 17 & 270\\
	    FIL8 & 19 & 300\\
	    FIL9 & 7 & 110\\
	    FIL10 & 9 & 150\\
	    FIL11 & 21 & 330\\
	    FIL12 & 15 & 240\\
	    FIL13 & 51 & 820 & 95 & 33\\
	    FIL14 & 19 & 300 & 107 & 29\\
	    FIL15 & 23 & 360 & 92 & 46\\
	    FIL16 & 11 & 180\\
	    FIL17 & 18 & 280\\
	    FIL18 & 21 & 330\\
	    FIL19 & 37 & 590\\
	    FIL20 & 6 & 100\\
	    FIL21 & 10 & 160\\
	    FIL22 & 17 & 270\\
	    FIL23 & 12 & 190\\
	    FIL24 & 11 & 180 & 81 & 37\\
	    FIL25 & 25 & 400 & 78 & 50\\
	    FIL26 & 13 & 200 & 84 & 40\\
	    FIL27 & 5 & 80 & 83 & 41\\
	    FIL28 & 6 & 100 & 79 & 33\\
            \noalign{\smallskip}
            \hline
	    \hline
	    \end{tabular}
     $$
\\
\footnotesize{NOTE: $\rm v_{exp}$ and the FWHM (corrected for instrumental
      broadening) are added when the ionized gas features were also detected in
      the spectra. The first part of the table contains the ring-like features
      which are divided into superbubbles (SB, with a diameter smaller than
      500\,pc) and supergiant shells (SGS, with a diameter larger than
      500\,pc). All other structures are presented in the second part of the
      table as filaments (FIL, all lengths). The same classification is also
      used for Table~\ref{Filsizeb}.} 
\end{table}
\begin{table}
\caption[]{The most prominent structures and their sizes in NGC\,4861.}
\label{Filsizeb}
     $$
         \begin{tabular}{lcccc}
	    \hline
	    \hline
	    \noalign{\smallskip}
	    Source & Diameter & Diameter & $\rm v_{helio}$ & FWHM\\
	    & [\arcs] & [pc] & [km/s] & [km/s]\\
	    \noalign{\smallskip}
	    \hline
	    \noalign{\smallskip}
	    SGS1 & 22 & 800\\
	    SGS2 & 20 & 730\\
            SGS3 & 24 & 890\\
	    SGS4$^{\mathrm{a}}$ & 30 & 1091 & 699/842 & 22/55\\
	    SB1 & 8 & 300 & 836 & 40\\
	    SB2 & 4 & 140 & 814 & 38\\
	    \noalign{\smallskip}
	    \hline
	    \hline
	    \noalign{\smallskip}
	    \noalign{\smallskip}
	    \noalign{\smallskip}
	    \hline
	    \hline
	    \noalign{\smallskip}
	    Source & Length & Length & $\rm v_{helio}$ & FWHM\\
	    & [\arcs] & [pc] & [km/s] & [km/s]\\
	    \noalign{\smallskip}
	    \hline
	    \noalign{\smallskip}
	    FIL1 & 49 & 1770\\
	    FIL2 & 5 & 170\\
	    FIL3 & 29 & 1070\\
	    FIL4 & 11 & 400 & 799 & 41\\
	    FIL5 & 18 & 640 & 823 & 31\\
	    FIL6 & 3 & 110\\
	    FIL7 & 2 & 70\\
	    FIL8 & 5 & 170\\
	    FIL9 & 12 & 440\\
	    FIL10 & 18 & 650\\
	    FIL11 & 26 & 940 & 842 & 36\\
	    FIL12 & 6 & 200\\
	    FIL13 & 4 & 130\\
	    FIL14 & 5 & 190\\
	    FIL15 & 13 & 470\\
	    FIL16 & 7 & 270\\
	    FIL17 & 7 & 240\\
	    FIL18 & 4 & 140\\
	    FIL19 & 5 & 180\\
	    FIL20 & 5 & 180\\
	    FIL21 & 5 & 170\\
	    FIL22 & 15 & 530\\
	    FIL23 & 17 & 620 & 793 & 40\\
	    FIL24 & 6 & 220\\
	    FIL25 & 13 & 480 & 751 & 36 \\
	    FIL26 & 9 & 330\\
	    FIL27 & 3 & 100\\
	    FIL28 & 2 & 70 & 797 & 28\\
	    FIL29 & 9 & 340 & 794 & 40\\
	    FIL30 & 6 & 200 & 797 & 36\\
	    FIL31 & 5 & 180\\
	    \noalign{\smallskip}
	    \hline
	    \hline
	    \end{tabular}
$$
\begin{list}{}{}
\item[$^{\mathrm{a}}$] This shell was not detected on the H$\alpha$ image of
  NGC\,4861 because the GEHR outshines all additional emission. The diameter
  of SGS4 was therefore estimated from the Doppler ellipse in slit~\emph{03}.
\end{list}
\end{table}
\newpage
\section{The echellograms}
\label{Appendix B}
\begin{figure*}[h]
   \centering
   \includegraphics[angle=90,width=8.1cm]{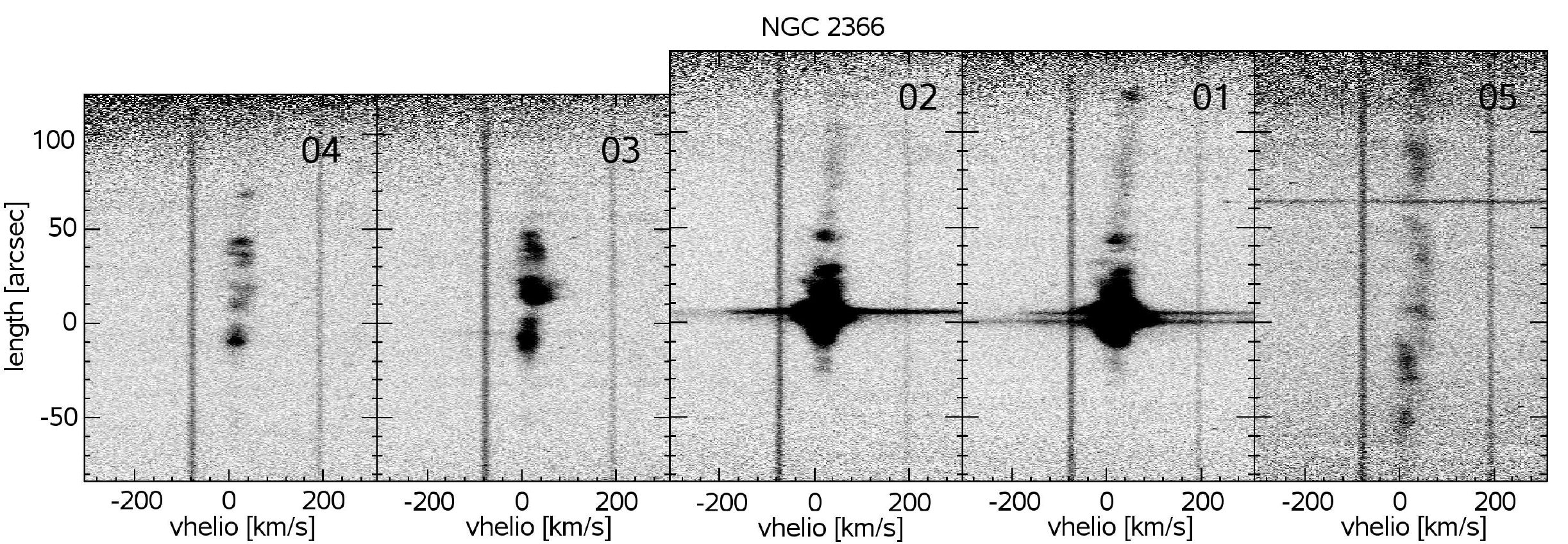}
   \hspace*{10mm}
   \includegraphics[angle=90,width=6.25cm]{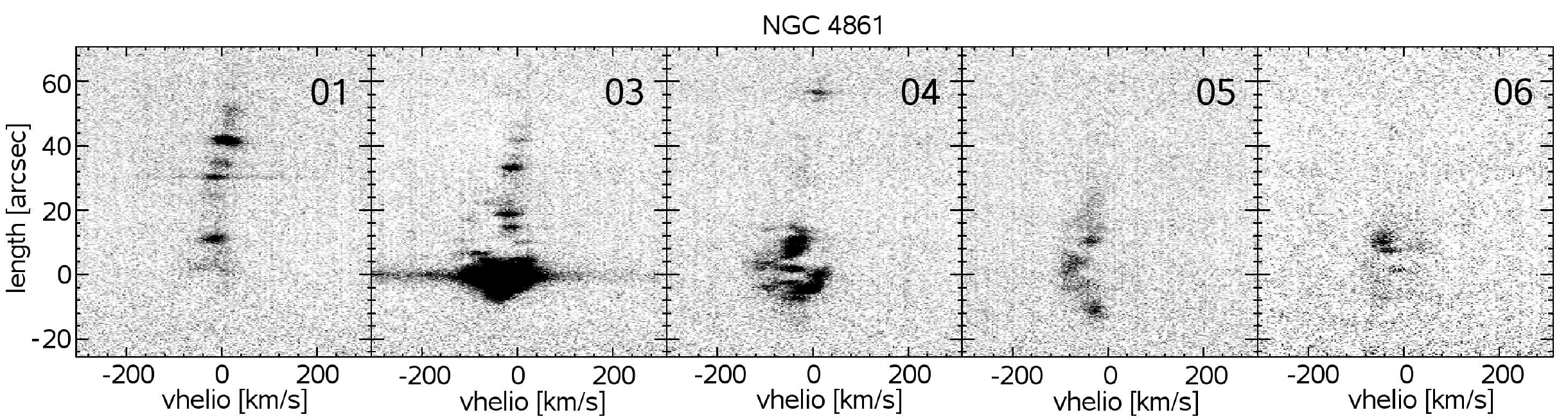}
      \caption{All echelle spectra of NGC\,2366 and NGC\,4861 arranged in a
      spatial sequence and centered on the
      H$\alpha$ line which is corrected for the systemic velocity of each
      galaxy.}
         \label{Figechspec}
   \end{figure*}
Figure~\ref{Figechspec} shows the echellograms obtained with the KPNO 4\,m
telescope and its echelle spectrograph. The position with respect to the
center of the GEHRs is drawn over the velocity with respect
to the H$\alpha$ line and corrected for the systemic velocity of the galaxies.
\end{document}